\documentclass[prl,twocolumn,preprintnumbers,amsmath,amssymb, superscriptaddress]{revtex4}

\usepackage{mathrsfs}
\usepackage{graphicx}
\usepackage{dcolumn}
\usepackage{bm}
\usepackage{color}

\newcommand{\figta}{(a)~}
\newcommand{\figtb}{(b)~}
\newcommand{\figtc}{(c)~}
\newcommand{\figa}{(a)}
\newcommand{\figb}{(b)}
\newcommand{\figc}{(c)}
\newcommand{\CgDsm}{\Delta s_\textrm{m}^\textrm{cc}}
\newcommand{\CgDstot}{\Delta s_\textrm{tot}^\textrm{cc}}
\newcommand{\Dstot}{\Delta s_\textrm{tot}}

\newcommand{\Qenv}{E_{E}}
\newcommand{\Benv}{\beta_{E}}

\newcommand{\be}{\begin{equation}}
\newcommand{\ee}{\end{equation}}
\newcommand{\bea}{\begin{eqnarray}}
\newcommand{\eea}{\end{eqnarray}}
\newcommand{\pf}{p_\rightarrow}
\newcommand{\pr}{p_\leftarrow}
\newcommand{\dpf}{\dot p_\rightarrow}
\newcommand{\dpr}{\dot p_\leftarrow}

\newcommand{\Rge}{\Gamma_{0 \rightarrow 1}}
\newcommand{\Reg}{\Gamma_{1 \rightarrow 0}}

\begin{document}

\title{Distribution of Entropy Production in a Single-Electron Box}

\author{J. V. Koski$^\ast$}
\affiliation{Low Temperature Laboratory (OVLL), Aalto University, POB 13500, FI-00076 AALTO, Finland}
\author{T. Sagawa}
\affiliation{Department of Basic Science, The University of Tokyo, Komaba 3-8-1, Meguro-ku, Tokyo 153-8902, Japan}
\author{O.-P. Saira}
\affiliation{Low Temperature Laboratory (OVLL), Aalto University, POB 13500, FI-00076 AALTO, Finland}
\affiliation{Kavli Institute of Nanoscience, Delft University of Technology, P.O. Box 5046, 2600 GA Delft, The Netherlands}
\author{Y. Yoon}
\affiliation{Low Temperature Laboratory (OVLL), Aalto University, POB 13500, FI-00076 AALTO, Finland}
\author{A. Kutvonen}
\affiliation{COMP Center of Excellence, Department of Applied Physics, Aalto University School of Science, P.O. Box 11000, FI-00076 Aalto, Espoo, Finland}
\author{P. Solinas}
\affiliation{Low Temperature Laboratory (OVLL), Aalto University, POB 13500, FI-00076 AALTO, Finland}
\affiliation{COMP Center of Excellence, Department of Applied Physics, Aalto University School of Science, P.O. Box 11000, FI-00076 Aalto, Espoo, Finland}
\author{M. M\"ott\"onen}
\affiliation{Low Temperature Laboratory (OVLL), Aalto University, POB 13500, FI-00076 AALTO, Finland}
\affiliation{COMP Center of Excellence, Department of Applied Physics, Aalto University School of Science, P.O. Box 13500, FI-00076 Aalto, Espoo, Finland}
\author{T. Ala-Nissila}
\affiliation{COMP Center of Excellence, Department of Applied Physics, Aalto University School of Science, P.O. Box 11000, FI-00076 Aalto, Espoo, Finland}
\affiliation{Department of Physics, Brown University, Providence RI 02912-1843, U.S.A.}
\author{J. P. Pekola}
\affiliation{Low Temperature Laboratory (OVLL), Aalto University, POB 13500, FI-00076 AALTO, Finland}

\date{\today}

\maketitle
\textbf{
Recently, the fundamental laws of thermodynamics have been reconsidered for small systems.  The discovery of the fluctuation relations~\cite{Evans,Gallavotti,Jarzynski1,Kurchan,Crooks} has spurred theoretical~\cite{Jarzynski2,Hatano,Seifert1,Sagawa,Kawai,Marin,Esposito1,SeifertR} and experimental~\cite{Wang,Liphardt,Trepagnier,Collin,Tietz,Blickle,Nakamura,Toyabe,Saira,Mehl2012} studies on thermodynamics of systems with few degrees of freedom.  The concept of entropy production has been extended to the microscopic level by considering stochastic trajectories of a system coupled to a heat bath. However, the experimental observation of the microscopic entropy production remains elusive. We measure distributions of the microscopic entropy production in a single-electron box consisting of two islands with a tunnel junction. The islands are coupled to separate heat baths at different temperatures, maintaining a steady thermal non-equilibrium.  As Jarzynski equality between work and free energy is not applicable in this case, the entropy production becomes the relevant parameter. We verify experimentally that the integral and detailed fluctuation relations are satisfied. Furthermore, the coarse-grained entropy production ~\cite{Kawai,Marin,Esposito1,Mehl2012,Kawaguchi} from trajectories of electronic transitions is related to the bare entropy production by a universal formula.  Our results reveal the fundamental roles of irreversible entropy production in non-equilibrium small systems.
}

Entropy production is a hallmark of irreversible thermodynamic processes. The concept of a stochastic microscopic trajectory allows one to define entropy for small systems \cite{Seifert1}. However, such trajectories depend on the scale of observation.  If one only accesses mesoscopic degrees of freedom, one observes coarse-grained trajectories of mesoscopic states.  The corresponding entropy production then differs from the bare entropy production without coarse-graining.  In fact,  it has been recently shown that coarse-graining of the slow background
degrees of freedom for stochastic dynamics may actually lead to a modification of the fluctuation relations for entropy \cite{Mehl2012}.
To clarify the concept of microscopic entropy production in non-equilibrium, accurate measurements are needed for systems, where the concepts
of stochastic dynamics and time scale separation between the system and the heat bath are well-defined. 

A single-electron box (SEB) device at low temperatures is an excellent test bench for thermodynamics in small systems \cite{Averin,Saira,Pekola2013}.

\begin{figure}[h!t]
\includegraphics[width=\columnwidth]{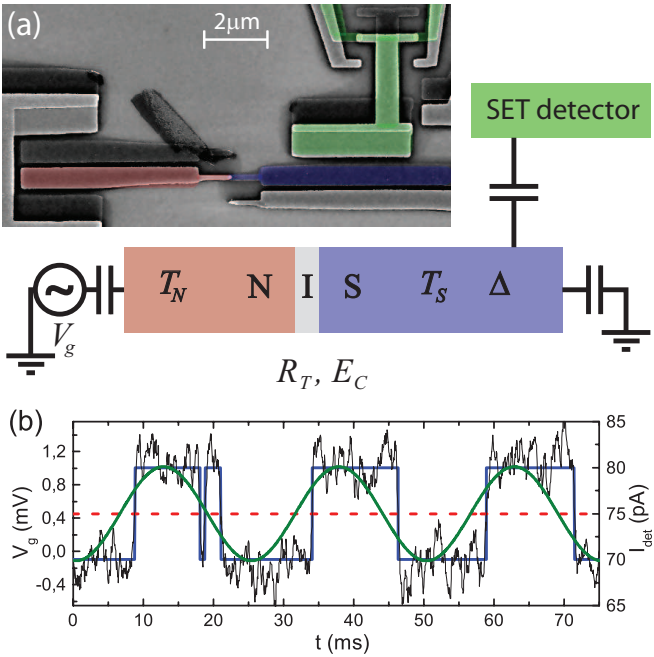}
\caption{\figta Sketch of the measured system together with a scanning electron micrograph of a typical sample. The colors on the micrograph indicate the correspondingly colored circuit elements in the sketch. The single electron box consists of a normal and a superconducting island that are connected through a tunnel junction. The charge state of the box is continuously probed with a SET detector. \figtb Example trace of the measurement data under sinusoidal protocol for the drive $V_g$, plotted in green. This trace covers three realizations of the forward process ($V_g$ from -0.1 to 1 mV), and three realizations of the backward process ($V_g$ from 1  to -0.1~mV). The SET current $I_{det}$, plotted in black, indicates the charge state of the box. The output of the threshold detection is shown in solid blue, with the threshold level as indicated by the dashed red line. The work and entropy production is evaluated for each realization separately, determined by the charge state trajectory.}
\label{fig:sample_design}
\end{figure}

The SEB employed here is shown in Fig.~\ref{fig:sample_design}\figa. The electrons in the normal-metal copper island (N) can tunnel to the 
superconducting Al island (S) through the aluminum oxide insulator (I). The sample fabrication~\cite{supp} methods are similar to those in Ref.~\cite{Saira}, 
but the design is different in that the S side of the junction does not overlap with the normal conductor in order to intentionally weaken the relaxation of energy in S~\cite{Maisi}. Moreover, the main results in Ref.~\cite{Saira} were extracted from measurements at the temperature of 220 mK, whereas these measurements are conducted at 140 mK. Lower temperature further weakens the relaxation significantly~\cite{Maisi}, leading to and elevated temperature in S. We denote by $n$ the integer net number of electrons tunneled from S to N relative to charge neutrality. As we can monitor the charge state $n$ with a nearby single-electron transistor (SET) shown in Fig.~\ref{fig:sample_design}\figa, we take our classical system degree of freedom to be $n$. 

The device in Fig.~\ref{fig:sample_design}(a) can be represented with a classical electric circuit, in which the energy stored in the capacitors 
and the voltage sources is given by~\cite{AL86,Averin,Pekola2}
\be \label{eq:CoulombPotential} H(n_g, n) = E_C (n - n_g)^2-e^2n_g^2/(2C_g), \ee
where $E_C$ is the characteristic charging energy, $C_g$ is the gate capacitance, $n_g=C_gV_g/e$ is the gate charge in units of the 
elementary charge $e$, and $V_g$ is the gate voltage which drives the system externally. 
Equation~\eqref{eq:CoulombPotential} gives the internal energy of the system.
In an instantaneous single-electron tunneling 
event from $n=k$ to $n=k+1$, the drive parameters stay constant and hence the work done to the system vanishes. 
Thus the first law of thermodynamics states that the generated heat is given by
\be\label{eq:heat} Q_k=H(n_g, k)-H(n_g, k+1)=E_C[2(n_g-k)-1]. \ee

It has been recently demonstrated that when the SEB is in thermal equilibrium and the transition rates obey the detailed balance condition, the Jarzynski
Equality (JE) $\langle e^{-\beta W}\rangle = e^{-\beta \Delta F}$ relating the work done in the system $W$ to its free energy change $\Delta F$
can be verified both theoretically and experimentally to a high degree of accuracy \cite{Averin,Pekola2,Saira,Pekola2013}.
In the present work, however, the two environments consisting of the excitations in the normal metal and the superconductor are at different 
temperatures $T_N=1/(k_B\beta_N)$ and $T_S=1/(k_B\beta_S)$, respectively, and hence the JE cannot be applied.
Nevertheless, we expect that our system should obey the so-called intergal fluctuation theorem~\cite{Seifert1}
\be \label{eq:SeifertEquality} \langle e^{-\Delta s_\textrm{tot}}\rangle = 1, \ee
for the total entropy production $\Delta s_\textrm{tot}=\Delta s + \Delta s_m$ given in terms of the increase of the 
system entropy $\Delta s = \ln\{P[n(t_f)]/P[n(0)]\}$ and the medium entropy production $\Delta s_m$. 
Here, $P[n(t)]$ is the directly measurable probability of the system to be in state $n$ at time instant $t$ 
given the initial condition and the drive $n_g$. We can express the entropy production of the medium as $\Delta s_m=\beta_N Q_N +\beta_SQ_S$, 
where $Q_N$ and $Q_S$ are the heat dissipated along the trajectory in the normal metal and in the superconductor, respectively.
We can measure the total dissipated heat $Q=Q_N+Q_S$ directly by monitoring $n(t)$ with the SET and using Eq.~\eqref{eq:heat}. 
The only essential assumption here is that the tunneling is elastic since the parameters of the Hamiltonian~\eqref{eq:CoulombPotential} 
can be measured independently. We can further obtain the conditional probability of $Q_N$ on $Q$ by some additional assumptions
(for technical details, see~\cite{supp}) and hence the probability distribution of $\Delta s_m$. 

On the other hand, medium entropy production can be defined by \cite{Seifert1,SeifertR}
\be \label{eq:MediumEntropy} \CgDsm = \sum_j \ln\left[\frac{\Gamma_{n_-\rightarrow n_+}(t_j)}{\Gamma_{n_+\rightarrow n_-}(t_j)}\right], \ee
where the system is taken to make transitions at time instants $t_j$ from the state $n_-$ to the state $n_+$, 
and $\Gamma_{n_-\rightarrow n_+}(t_j)$ and $\Gamma_{n_+\rightarrow n_-}(t_j)$ are the corresponding forward and backward transition rates. 
We refer to $\Delta s_m^\textrm{cc}$ as the {\it coarse-grained (cc) medium entropy production} 
since it can be shown analytically \cite{supp} that in a single tunneling event
\be \langle e^{-\Delta s_m} \rangle_Q = e^{-\CgDsm(Q)}, \ee
where the average is taken over $Q_N$ for a fixed $Q$.
This equality further implies $\langle \Delta s_m \rangle_Q \geq \Delta s_m^\textrm{cc}(Q)$ and provides 
a physical interpretation for $\Delta s_m^\textrm{cc}$, the definition of which in Eq.~\eqref{eq:MediumEntropy} 
coincides with the definition of medium entropy for general stochastic systems~\cite{SeifertR}. 
Note that by introducing transition rates, we have implicitly assumed that the system is Markovian, 
a fact that can be experimentally verified in our setup.

As mentioned above, we can obtain experimentally the probability distributions 
$P_{\rightleftarrows}(\Delta s_\textrm{tot})$ and $P_{\rightleftarrows}(\Delta s_\textrm{tot}^\textrm{cc})$ 
of $\Delta s_\textrm{tot}=\Delta s + \Delta s_m$ and $\Delta s_\textrm{tot}^\textrm{cc}=\Delta s + \Delta s_m^\textrm{cc}$, respectively, 
and hence access the integral fluctuation theorem of Eq.~\eqref{eq:SeifertEquality} that should be satisfied by all the distributions. 
Here, $P_\rightarrow$ is the distribution for a forward driving protocol $n_{g,\rightarrow}(t)$ and $P_\leftarrow$ corresponds to the 
backward protocol $n_{g,\leftarrow}(t)=n_{g,\rightarrow}(t_f-t)$. In addition, we expect our system to satisfy so-called detailed fluctuation relations~\cite{Crooks, SeifertR}
\bea \label{eq:DetailedFluctuation} 
P_{\rightleftarrows}(\CgDstot) / P_{\leftrightarrows}(-\CgDstot) &=& e^{\CgDstot}, \\
P_{\rightleftarrows}(\Delta s_\textrm{tot}) / P_{\rightleftarrows}(- \Delta s_\textrm{tot}) &=& e^{\Delta s_\textrm{tot}}.\nonumber
\eea

In our experiments, we drive the system with the gate charge $n_g(t) = n_0 - A \cos(\pi f t)$, where $n_0 \approx A \approx 0.5$. 
Figure~\ref{fig:sample_design}(b) shows the applied drive and an example trace of the detector current. 
Clearly, two discrete current levels corresponding to the charge states $n=0$ and $n=1$ are observable. 
Due to the low bath temperatures, $130-160$ mK, the relatively high charging energy $E_C\approx 162$~$\mu$eV~$=1.88~\textrm{K}\times k_B$, 
and low driving frequencies $f\leq 120$ Hz, the system essentially always finds the minimum-energy state at the extrema of the drive. 
Thus we can partition the continuous measurement into legs of forward and backward protocols, 
for which the charge state and gate charge change from $0$ to $1$ and $1$ to $0$, respectively. 
Conversion of the current trace from such a leg using threshold detection yields a realization for a 
system trajectory $n(t)$ which is used in the ensemble average with unit weight to obtain the desired distributions. 
The charging energy, the temperatures of the normal metal and the superconductor, the tunneling resistance of the junction 
$R_T\approx 1.7~\textrm{ M}\Omega$, and the excitation gap of the superconductor $\Delta\approx 224$~$\mu$eV are determined experimentally~\cite{supp}.

\begin{figure} [h!b]
\includegraphics[width=\columnwidth]{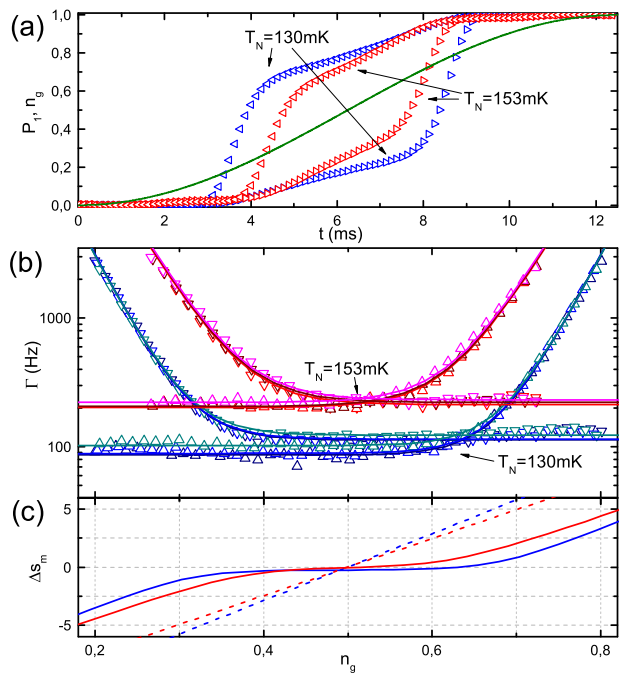}
\caption{\figta Probability for $n=1$ as a function of time, extracted from the measured ensemble of processes with drive frequency 40 Hz. The triangles pointing right indicate the probability for a forward process, and those pointing left indicate the probability for a backward process. The sinusoidal drive $n_g$ is displayed as a solid green line. \figtb The tunneling rates obtained from measured probabilities \cite{supp} (shown in triangles) for measurements $1-3$ and $8-10$ listed in Table \ref{tab:Parameters} 
with their corresponding fits (solid lines). 
\figtc $\Delta s^{\textrm{cc}}_m$ of Eq. (\ref{eq:MediumEntropy}) for the transition 0 $\rightarrow$ 1 as a function of $n_g$, obtained from the tunneling rates in \figb. The dashed lines show $\beta_N Q$, which is equivalent to both $\Delta s^{\textrm{cc}}_m$ and $\Delta s_m$ if $T_S = T_N$.}
\label{fig:Rates}
\end{figure}

Since the charge state corresponds to the ground state in the beginning and at the end of the drive, 
the system entropy change $\Delta s$ in Eq.~\eqref{eq:SeifertEquality} and the free-energy change 
vanish, and we thus only need to obtain $\Delta s_m$ 
in order to assess if the fluctuation relations are satisfied.
To determine $\Delta s_m^\textrm{cc}$ in Eq. (\ref{eq:MediumEntropy}), the time-dependent tunneling 
rates $\Gamma_{i\rightarrow j}(t)$ need to be measured. 
To extract them from an ensemble of forward and backward pumping trajectories, we show in Fig. \ref{fig:Rates}(a) the observed 
probabilities for the system charge state to be $n=1$ for the forward and backward drives denoted by $P^1_\rightleftarrows$. 
The rates can be solved by comparing the measured data to the outcome from the master equation~\cite{supp}.
An example of results obtained this way are shown in Fig.~\ref{fig:Rates}(b). 
The rates from the standard sequential tunneling model~\cite{supp} are in agreement with the experimentally obtained data. Here $T_N$ is assumed to be the temperature of the cryostat, while $T_S$ is obtained for each measurement from the fit as listed in Table 1.

The medium entropy production for a tunneling event with coarse graining, $\CgDsm(n_g)$, 
extracted from the fitted rates is shown in Fig.~\ref{fig:Rates} \figc, demonstrating the significant effect of the overheating of the superconductor. 
For $n_g \approx 0.4 - 0.6$, the tunneling probability is primarily determined by the thermal excitations of the superconductor and not by $n_g$. 
Thus, the rates for different directions are almost equal and $\Delta s_m^\textrm{cc}$ is nearly vanishing. 
Only tunneling events that occur outside this $n_g$ range contribute significantly to the cumulative entropy production.

\begin{table}[h!t]
\caption{Measurement parameters and obtained averages for work and entropy production. $T_{S,0/1}$ is the S temperature matching the state $n = 0/1$, and $\bar W = W - \Delta F$ } \label{tab:Parameters}
\centering
\begin{tabular}{c c c c c c c c c} 
\hline\hline \\[0.5ex]
Meas. & $f$& $T_N$& $n_0$ & $T_{S, 0}$ & $T_{S, 1}$ & $\langle e^{-\beta_N \bar W}\rangle$ & $\langle e^{-\CgDstot}\rangle$ & $\langle e^{-\Dstot}\rangle$\\ 
& $\textrm{(Hz)}$& $\textrm{(mK)}$& & $\textrm{(mK)}$ & $\textrm{(mK)}$ & &  \\
[0.5ex] 
\hline
1 & 20 & 130 & 0.526 & 174 & 177 & 93 & 1.085 & 1.063\\
2 & 40 & 130 & 0.516 & 174 & 177 & 129 & 1.064 & 1.053\\
3 & 80 & 130 & 0.507 &176 & 178 & 180 & 1.074 & 1.083\\
4 & 20 & 142 & 0.513 &179 & 181 & 20 & 1.064  & 1.030\\
5 & 40 & 142 & 0.509 &179 & 181 & 30 & 1.054  & 1.047\\
6 & 80 & 142 & 0.505 &180 & 181 & 45 & 1.096  & 1.100\\
7 & 120 & 141 & 0.504 &181 & 182 & 68 & 1.241  & 1.324\\
8 & 40 & 153 & 0.502 &184 & 184 & 11 & 1.095 & 1.058\\
9 & 80 & 153 & 0.503 &184 & 185 & 15 & 1.140 & 1.139\\
10 & 120 & 153 & 0.502 &185 & 186 & 20 & 1.301 & 1.370\\ [0.5ex] 
\hline
\end{tabular}
\label{table:nonlin} 
\end{table}

\begin{figure}[h!t]
\includegraphics[width=0.98\columnwidth]{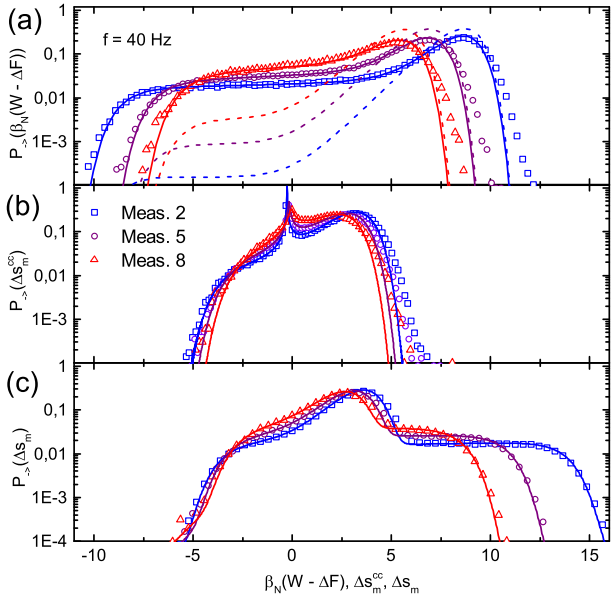}
\caption{
\figta $P_\rightarrow(\beta_N (W - \Delta F))$ distributions for 40 Hz forward process at different bath temperatures. The symbols show measured values (all panels), solid lines are numerical expectations (all panels), and dashed lines demonstrate what the distribution would be for $T_S = T_N$. 
\figtb Corresponding $P_\rightarrow(\CgDstot)$ distributions.
\figtc $P_\rightarrow(\Dstot)$ distributions for single jump trajectories. }
\label{fig:WorkAndEntropyDistributions}
\end{figure}
\begin{figure}[h!b]
\includegraphics[width=0.98\columnwidth]{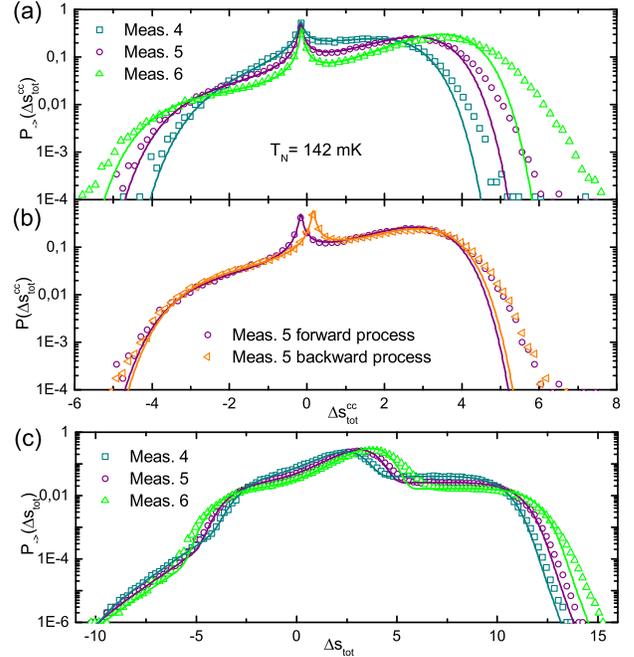}
\caption{\figta $\CgDstot$ distributions obtained at $T_N = 142$ mK with different drive frequencies. \figtb Distributions of an individual measurement for forward and backward processes. \figtc Bare entropy distributions for the measurements at $T_N = 142$ mK. }
\label{fig:FrequencyDistributions}
\end{figure}

\begin{figure} [h!b]
\includegraphics[width=\columnwidth]{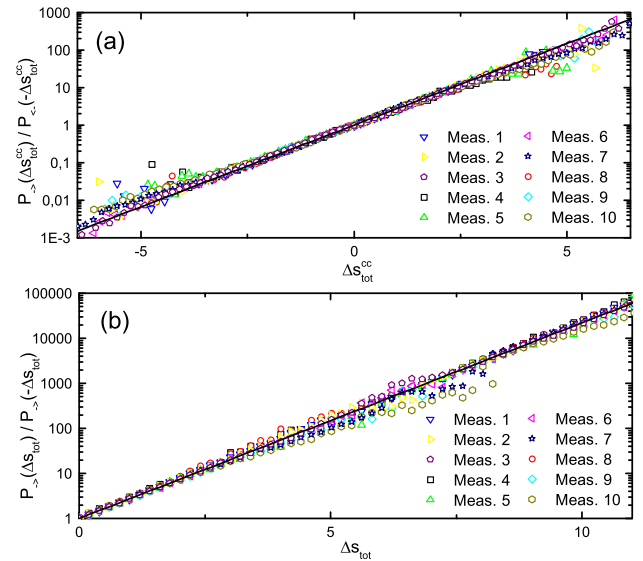}
\caption{\figta Test of the detailed fluctuation relation, Eq. (\ref{eq:DetailedFluctuation}), for the measured trajectory entropy distributions. Despite the asymmetry of forward and backward processes due to detector back-action, the relation is satisfied. \figtb The detailed fluctuation relation for bare entropy production of the forward processes.  The solid lines in (a) and (b) show the expected dependence given by Eq. (\ref{eq:DetailedFluctuation}). }
\label{fig:DetailedFluctuation}
\end{figure}

Table \ref{tab:Parameters} presents a collection of exponential averages for entropy production as to test the integral fluctuation relation. 
Figure~\ref{fig:WorkAndEntropyDistributions} shows the experimentally obtained distributions for work and entropy production 
with coarse graining together with the theoretical predictions for $f = 40$~Hz and different $T_N$. 
For comparison, the prediction of JE at the bath temperature is shown by the dashed lines.
As expected, the measured work distributions in Fig.~\ref{fig:WorkAndEntropyDistributions}(a) do not follow from JE, which would assume just one temperature. 
The difference between $T_N$ and $T_S$ decreases with increasing $T_N$, and hence the difference between data and the dashed lines decreases as well. 
All the entropy distributions in Fig.~\ref{fig:WorkAndEntropyDistributions}(b), obtained from the same trajectories as the work distributions, 
satisfy the integral fluctuation theorem within the errors. The peaks in the distributions in the vicinity of zero arise from the 
relatively long time spent during the drive in the region $n_g \approx 0.4 - 0.6$, where the entropy production is nearly vanishing, see Fig.~\ref{fig:Rates}(c).

Figure~\ref{fig:FrequencyDistributions} \figta displays the probability distributions of entropy production at fixed $T_N = 142$~mK for various frequencies. 
The tails of the distribution broaden and the peak at $\CgDstot =0$ sharpens with increasing frequency. 
Figure \ref{fig:FrequencyDistributions} \figtb shows the distributions for forward and backward processes. 
The distributions are overlapping, apart from the positions of the peaks near vanishing $\CgDstot$. 
The offset of the peaks is explained by different superconductor temperatures for different tunneling directions leading to 
$\Gamma_{0\rightarrow 1}(n_g = 0.5) < \Gamma_{1\rightarrow 0}(n_g = 0.5)$, and hence an offset of the point 
$\CgDstot = 0$ away from $n_g=0.5$. Because the charge degeneracy point is the most probable 
point for the tunneling to take place, the peak is located there. For the $n:0\to 1$ event, this corresponds to 
negative entropy production and positive production is observed for $n:1\to 0$. Different temperatures for 
different tunneling directions can be justified by the difference in the observed tunneling rates in 
Fig.~\ref{fig:Rates}\figb. The SET current is higher for $n=1$ than for $n=0$ 
[see Fig.~\ref{fig:sample_design}(c)], inducing a higher excess heating power for the superconductor at $n=1$. However, even with the offset in the distributions in Fig.~\ref{fig:FrequencyDistributions}(b), they obey the detailed fluctuation theorem, as shown in Fig. \ref{fig:DetailedFluctuation}.

To summarize, we have extracted the distributions of work, bare entropy production, and coarse-grained entropy production for a sinusoidal drive protocol in a single-electron box. Due to the thermal non-equilibrium caused by the overheating of the S island, the work and entropy distributions no longer coincide. As a consequence, the Jarzynski equality is no longer applicable, but the integral and detailed fluctuation relations for entropy production are shown to be valid. We also find that the two different measures of entropy are related by a universal formula.

This work has been supported in part by Academy of Finland though its LTQ (project no. 250280) and COMP (project no. 251748) CoE grants, 
the Research Foundation of Helsinki University of Technology, and V\"ais\"al\"a Foundation. We acknowledge Micronova Nanofabrication Centre of Aalto University for providing the processing facilities and technical support. 
We thank Ville Maisi, Frank Hekking, and Simone Gasparinetti for useful discussions.

\clearpage
\section{\Large Supplementary Material}

\section{Relation between bare and coarse grained entropy}

In the following, we derive the equality 
\be \label{eq:entropyAverage} \langle e^{-\Delta s_m}\rangle = e^{-\Delta s_m^\textrm{cc}} \ee
 for trajectories with a single tunneling event in a  NIS single electron box. Let $n$ be the net number of electrons tunneled from $S$ to $N$. An electron may either tunnel from the S island to the N island, or tunnel from N-lead to the S-lead, adding or substracting one to the parameter $n$. Let the tunneling events of the former type be denoted with $'+'$, and of the latter with $'-'$. We consider a box at an electromagnetic environment at temperature $T_E$. Upon a tunneling event, the environment absorbs or emits a photon with energy $E_{E}$: the energy of the electron in the superconducting lead $E_S$ and the energy in the normal lead $E_N$ then satisfy
\be E_S - E_N = \pm(Q + \Qenv), \ee
where $Q$ is the dissipated heat upon the tunneling process $0\rightarrow 1$, determined solely by the external control parameter $n_g$. The dimensionless medium entropy change of such an event is
\be \Delta s_{m}^\pm = \mp \beta_S E_S \pm \beta_N E_N \pm \beta_{E} \Qenv, \ee
where $\beta_i = 1 / k_B T_i$ denotes the inverse temperature.

According to the theory for sequential tunneling in an electromagnetic environment
\cite{Ingold}, the tunneling rates for transitions $\Rge \equiv \Gamma_+$ and $\Reg \equiv \Gamma_-$ are
\be \label{eq:Trate} \begin{split}
\Gamma_\pm(Q) =& \int dE_S \int d\Qenv \gamma_\pm(E_S, Q, \Qenv); \\
\gamma_\pm(E_S, Q, \Qenv) =& \frac{1}{e^2R_T} N_S(E_S) f_S(\pm E_S) \times \\
& P(\pm\Qenv) f_N(\mp E_N),
\end{split}\ee
where $R_T$ is the tunneling resistance, $N_S(E) = \textrm{Re} \left(|E| / \sqrt{(E^2 - \Delta^2)}\right)$ is the normalized BCS superconductor density of states with a superconductor energy gap $\Delta$, $P(\Qenv)$ is the probability for the environment to absorb the energy $Q_E$, and $f_{N/S}(E_{N/S}) = (1 + \exp(\beta_{N/S} E_{N/S}))^{-1}$ is the fermi function of the N/S lead, giving the probability for an electron to occupy the energy level $E_{N/S}$. The conditional probability for the energy parameters to be exactly $E_S$ and $\Qenv$ is $P(E_S, \Qenv~|~Q, n\rightarrow n \pm 1) = \gamma_\pm(E_S, Q, \Qenv) / \Gamma_\pm(Q).$

Left hand side of Eq. (\ref{eq:entropyAverage}) becomes
\be \langle e^{-\Delta s_{m}^\pm}\rangle =  \int dE_S \int d\Qenv e^{-\Delta s_{m}^\pm} \frac {\gamma_\pm(E_S, Q, \Qenv)}{\Gamma_\pm(Q)}. \ee
Since the environment function satisfies detailed balance, $P(\Qenv) / P(-\Qenv) = e^{\Benv \Qenv}$, and the fermi function satisfies $e^{\beta_{N/S} E_{N/S}} f_{N/S}(E_{N/S}) = f_{N/S}(-E_{N/S})$, one obtains $e^{-\Delta s_{m}^\pm} \gamma_\pm(E_S, Q, \Qenv) = \gamma_\mp(E_S, Q, \Qenv)$, and with Eq. (\ref{eq:Trate}) the average is then
\be \langle  e^{-\Delta s_{m}^\pm}\rangle = \frac{\Gamma_\mp(Q)}{\Gamma_\pm(Q)} = e^{-\Delta s_m^{\pm,\textrm{cc}}}\ee

\section{Extraction of tunneling rates from state probabilities}
The tunneling rates $\Reg$ and $\Rge$ are obtained from the master equation for a two state system. At any given time instant $t$, the system has a probability $P_1$ to occupy the charge state $n = 1$. As the charge state must be either $n = 0$ or $n = 1$, the occupation probability for $n = 0$ is $P_0 = 1 - P_1$. The master equation is then
\be \dot P_1 = -\Reg(n_g(t)) P_1 + \Rge(n_g(t))(1 - P_1). \ee
In order to solve the tunneling rates as a function of $n_g$, the occupation probability is calculated for both forward $n_g^\rightarrow(t)$ and reverse drives $n_g^\leftarrow(t)$. These satisfy $n_g^\leftarrow(t) = n_g^\rightarrow(\tau - t)$, and by a change of variable $t' = \tau - t$, $n_g^\leftarrow(t) = n_g^\rightarrow(t')$, two equations are obtained:
\be 
\begin{split}
\dot P_1^\rightarrow &= -\Reg(n_g^\rightarrow(t)) P_1^\rightarrow + \Rge(n_g^\rightarrow(t))(1 - P_1^\rightarrow); \\ 
-\dot P_1^\leftarrow &= -\Reg(n_g^\rightarrow(t')) P_1^\leftarrow + \Rge(n_g^\rightarrow(t'))(1 - P_1^\leftarrow). 
\end{split}
\ee
The rates are then solved as
\be \label{eq:RateEquations} \begin{split}
\Reg(n_g^\rightarrow(t)) &= \frac{\dpr (1 - \pf) + \dpf(1 - \pr)}{\pr - \pf};\\
\Rge(n_g^\rightarrow(t)) &= \frac{\dpf\pr + \dpr \pf}{\pr - \pf}. \\
\end{split}\ee

$\pf$ and $\pr$ are obtained from the measurements for a time interval $t... t + \Delta t$ by averaging $n$ over the ensemble of process repetitions. The obtained distributions are inserted in Eq. (\ref{eq:RateEquations}) to obtain the rates. As shown in Fig. 2 (b), Eq. (\ref{eq:Trate}) describes the extracted tunneling rates well. The direct effect of the environment is negligible, and we take the limit of weak environment, $P(\Qenv) = \delta(\Qenv)$.
By fitting  Eq. (\ref{eq:Trate}) to the measured rates, we obtain $R_T \simeq 1.7~$M$\Omega$, $\Delta \simeq 224~\mu$eV, and $E_C \simeq 162~\mu$eV. Here, $T_N$ is assumed to be the temperature of the cryostat, while $T_S$ is obtained for each measurement separately as listed in Table 1.

\section{Fabrication methods}
The sample was fabricated by the standard shadow evaporation technique \cite{77-Dolan}. The superconducting structures are aluminium with a thickness of $\simeq 25$ nm. The tunnel barriers are formed by exposing the aluminium to oxygen, oxidizing its surface into an insulating aluminium oxide layer. The normal metal is copper with a thickness of $\simeq 30$ nm.

\end{document}